\documentclass[cameraready]{Interspeech}

\title{FlowTTS-GRPO: Online Reinforcement Learning with Multi-Objective Reward Optimization for Flow-Matching Based Text-to-Speech}

\author[orcid=0009-0000-0128-5421]{Haoxu}{Wang}
\author[]{Biao}{Tian}
\author[orcid=0009-0001-0317-727X]{Weiqin}{Li}
\author[orcid=0000-0002-8866-6637]{Xiang}{Lv}
\author[]{Han}{Zhao}
\author[]{Xiangang}{Li}

\address{
    Tongyi Lab, Alibaba Group, China
}

\email{\{wanghaoxu.whx, tianbiao.tb, liweiqin.lwq, lyuxiang.lx, quantu.zh, lixiangang.lxg\}@alibaba-inc.com}


\usepackage{hyperref}
\usepackage{comment}

\usepackage{cite}
\usepackage{amsmath,amssymb,amsfonts}
\usepackage{algorithmic}
\usepackage{graphicx}
\usepackage{textcomp}
\usepackage{xcolor}
\usepackage{multirow}
\usepackage{booktabs}
\usepackage{marvosym}
\usepackage{cite}
\usepackage{setspace} 
\usepackage{makecell}
\usepackage[table]{xcolor}

\usepackage{pifont} 

\usepackage{CJKutf8} 

\begin{document}

\maketitle

\begin{abstract}

Existing Reinforcement Learning (RL) research for Text-to-Speech (TTS) focuses on large language models (LLMs), leaving Flow-Matching (FM) under-explored. We present FlowTTS-GRPO, an online RL framework for FM-based TTS. By converting ordinary differential equation (ODE) trajectories into stochastic differential equation (SDE) paths, our method enables direct fine-tuning of open-source FM models without auxiliary models. We show that a weighted reward combination converges faster than a probabilistic scheme, and identify three practical optimizations: omitting classifier-free guidance (CFG) during training accelerates convergence; synthesizing hard cases improves robustness; and applying RL to the FM component enhances audio-detail metrics. Experiments on CosyVoice 3.0 and F5-TTS demonstrate objective and subjective preference gains in speaker similarity and perceptual quality, with F5-TTS also improving intelligibility.
\end{abstract}

\section{Introduction}
\label{sec:intro}

Text-to-speech (TTS) converts input text into audible speech and plays a key role in human–computer interaction. Recently, researchers have incorporated large language models (LLMs)\cite{kaplan2020scaling} into TTS\cite{wang2023neural,du2023lauragpt,du2024cosyvoice1,du2024cosyvoice2,du2025cosyvoice3,ye2025llasa,wang2025sparktts,anastassiou2024seedtts}. One line of work uses LLMs' ability to model discrete speech tokens with in-context learning (ICL)\cite{wang2023neural,ye2025llasa,wang2025sparktts}: a speech codec\cite{zeghidour2021soundstream,defossez2022high_encodec} produces discrete tokens, an LLM autoregressively (AR) predicts these tokens, and a decoder reconstructs the waveform. This pipeline is constrained by the codec's capacity and requires a trade-off between semantic tokens and acoustic tokens\cite{guo2025recent_tokensreview}. Another approach\cite{chen2025f5tts} relies solely on non‑autoregressive (NAR) Flow Matching (FM)\cite{flowmatching}: compared with token-decode pipelines it produces finer acoustic details and offers faster inference than AR models, but struggles to incorporate semantic information, predict speech duration accurately, and support streaming. A hybrid paradigm has emerged\cite{du2024cosyvoice1,du2024cosyvoice2,du2025cosyvoice3}, where LLMs model semantic tokens and FM recovers acoustic detail. This scheme captures text–token semantics while enhancing acoustic fidelity and naturalness, outperforming earlier paradigms with higher synthesis success rates\cite{du2025cosyvoice3,zhou2025indextts2}.

Although LLM–FM hybrid TTS models can generate diverse speech, some outputs remain difficult to align with human perceptual preferences. Meanwhile, post‑training aligns models to human perception or downstream task objectives and is applied across many generative tasks (language understanding\cite{ouyang2022training}, image generation\cite{xu2023imagereward}), further fine‑tuning a pretrained generative model using reinforcement learning (RL) to improve human‑perception metrics.

\begin{figure}[!t]
	\centering
\includegraphics[width=1.0\linewidth]{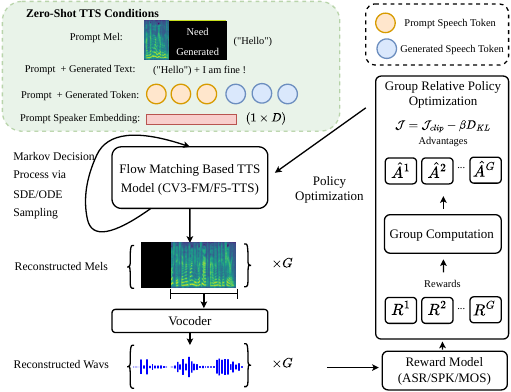}
	\caption{
        The pipeline of our FlowTTS-GRPO post-training for CosyVoice 3.0 and F5-TTS. Prompt speech tokens, generated speech tokens, and prompt speaker embedding are only used for CosyVoice 3.0.
	}
	\label{fig:grpo}
\end{figure}

In the TTS domain, RL work has largely focused on LLM‑based models. Seed‑TTS\cite{anastassiou2024seedtts} applies proximal policy optimization (PPO)\cite{schulman2017proximal_ppo} to optimize Word Error Rate (WER) and speaker similarity (SS), but PPO requires training multiple auxiliary models, is complex and unstable, and yields noticeable improvement mainly in WER. Other works\cite{tian2025preference_ttsdpo,gao2025emodpo,hussain2025koel} use Direct Preference Optimization (DPO)\cite{dpo} and its variants to strengthen LLMs; although they avoid maintaining extra models during training, they demand large numbers of human‑preference pairs, incurring heavy inference costs and expensive annotations. DiffRO\cite{gao25d_interspeech} directly applies differentiable reward optimization on speech tokens to improve intelligibility and expressiveness, using Gumbel‑Softmax to optimize an Automatic Speech Recognition (ASR) loss; however, it still requires a separately trained token‑to‑reward model and is sensitive to reward‑model bias, which RRPO\cite{wang2025rrpo} addresses by training a more robust reward model. \cite{liu2025group_llmttsgrpo} fine‑tunes LLM‑based TTS with Group Relative Policy Optimization (GRPO) and a compound CER+NLL reward. Crucially, these studies concentrate on LLMs and there is little RL research for FM–based TTS. F5R‑TTS\cite{sun2025f5r} is the first work to apply RL on FM–based TTS. It makes preliminary RL gains in WER and SS but needs an independently trained Gaussian FM generator to produce randomness and cannot leverage widely used open-source models; moreover, it uses a single rollout per prompt (rollout=1), failing to exploit multiple similar syntheses per prompt, group-wise advantage gaps, and multi‑objective optimization.

Flow‑GRPO\cite{flowgrpo} is the first method to introduce online reinforcement learning into flow‑matching models. It transforms the Ordinary Differential Equation (ODE) trajectory into a Stochastic Differential Equation (SDE), supplying the stochasticity required by RL and thereby improving human‑perception metrics in image synthesis. FlowSE‑GRPO\cite{wang2026flowse} adapts this technique to speech enhancement. Unlike image synthesis or speech enhancement, zero-shot TTS must jointly preserve speaker identity, linguistic content, and perceptual quality under text and prompt-speech conditioning. This creates reward conflicts and architecture-specific behavior that have not been studied in prior Flow-GRPO applications.

In this paper, we propose FlowTTS-GRPO, an online RL framework for FM-based TTS. Our contributions are: (1) We present the first successful introduction of Flow-GRPO to TTS by converting ODE trajectories to SDE paths, enabling direct fine-tuning of open-source FM models without retraining a separate stochastic generator. (2) FlowTTS-GRPO simplifies prior RL approaches by eliminating auxiliary models (value networks, preference pairs, token-to-reward models) and leveraging off-the-shelf reward signals. (3) We analyze multi-objective reward conflicts in TTS and find that weighted combination with standard deviation normalization yields faster and more stable convergence than probabilistic combination. (4) We identify three practical optimizations and observations: omitting classifier-free guidance (CFG) during training accelerates convergence, hard case synthesis enhances robustness and accelerates RL training, and RL on FM primarily improves audio-detail metrics while RL on LM targets intelligibility for LLM-FM hybrid models. Experiments on CosyVoice 3.0\cite{du2025cosyvoice3} (LLM-FM hybrid, where intelligibility is governed by the LM and FM controls acoustic details) and F5‑TTS\cite{chen2025f5tts} (FM-only, where FM can improve both intelligibility and audio details) demonstrate significant improvements in speaker similarity and perceptual quality, with F5‑TTS also achieving reduced WER, and robust generalization across speaker verification models, languages, and LLM front-ends.

\section{Methods}
\label{sec:method}

\subsection{The TTS model used for RL finetuning}

We select CosyVoice 3.0\cite{du2025cosyvoice3} (CV3) and F5‑TTS\cite{chen2025f5tts} as pretrained TTS models for RL finetuning. Zero-shot voice-cloning refers to generating speech in a target speaker's voice using only a short prompt audio without requiring explicit speaker adaptation or fine-tuning. To refine zero-shot voice-cloning behavior, we optimize the model with respect to its actual inference procedure. We treat a training utterance as the prompt waveform and the prompt text, randomly sample a new target text, run the model's inference to synthesize a generated waveform, and use the resulting (prompt, generated) pair as the RL training sample.

\subsubsection{CosyVoice 3.0}

CosyVoice 3.0 is a widely used TTS system that bridges an LLM with FM; its pipeline includes a speech tokenizer, an LM, an FM acoustic model, and a vocoder. At inference, the speech tokenizer extracts prompt tokens from the prompt wav. The LM autoregressively generates speech tokens conditioned on the prompt text and the target text. The FM model, conditioned on the concatenation of prompt and generated tokens plus the prompt mel and prompt speaker embedding, produces the generated mel. Finally, the vocoder\cite{hifigan} converts the generated mel into the generated wav, yielding the cloned voice. In this paper, we optimize only the FM component and do not fine-tune the speech tokenizer, LM, or vocoder.

\subsubsection{F5-TTS}

F5-TTS is an FM-only TTS system consisting of an FM acoustic model and a vocoder. At inference, the FM model conditions on the concatenation of prompt text and generated text together with the prompt mel to produce the generated mel. The vocoder\cite{siuzdak2023vocos} then converts the generated mel to the generated wav to produce the cloned audio. In this paper, we fine-tune only the FM component and keep the vocoder fixed.

\subsection{FlowTTS-GRPO}

\subsubsection{Markov Decision Process}

The FM-based TTS model $\phi_{\theta}$ predicts velocity $v = x_{1} - x_{0}$ from $x_{t} = (1 - t)x_{0} + t x_{1}$, where $x_{1} \sim X_{1}$ is real speech and $x_{0} \sim X_{0} = \mathcal{N}(0,\mathbf{I})$.

Following FlowSE-GRPO\cite{wang2026flowse}, the FM decoding process is formulated as a Markov decision process $(S, A, \rho_{0}, P, R)$ where time $t$ evolves from $0$ to $1$ over $T$ discrete steps (e.g. step size $\Delta t = 1/T$). The components are defined as follows:

\begin{itemize}
    \item \textbf{State} $s_{t} \in S$: $s_{t} \triangleq (c, t, x_{t})$ with conditionals $c$ (e.g., text phonemes, speaker embeddings) and latent $x_{t}$.
    
    \item \textbf{Action} $a_{t} \in A$: Predicted velocity $a_{t} \triangleq v_{t} = \phi_{\theta}(x_{t}, c)$ with policy $\pi(a_{t} \mid s_{t}) = p_{\theta}(v_{t} \mid x_{t}, c)$.
    
    \item \textbf{Initial State Distribution} $\rho_{0}$: $\rho_{0}(s_{0}) \triangleq (p_{c}, \delta_{0}, \mathcal{N}(0, \mathbf{I}))$, starting from Gaussian noise at $t=0$.
    
    \item \textbf{Transition Dynamics} $P$: Deterministic Euler update. Given $s_t$ and $a_t$, the next state is:
    \begin{equation}
        p(s_{t+\Delta t} \mid s_{t}, a_{t}) \triangleq (\delta_{c}, \delta_{t+\Delta t}, \delta_{x_{t} + v_{t} \Delta t})
    \end{equation}
    with recursive update $x_{t+\Delta t} = x_{t} + v_{t} \Delta t$, where $\delta_{y}$ denotes a Dirac delta distribution centered at $y$.
    
    \item \textbf{Reward} $R$: Terminal reward at $t=1$:
    \begin{equation}
        R(s_{t}, a_{t}) \triangleq 
        \begin{cases} 
        r(x_{1}, c) & \text{if } t = 1 \\
        0 & \text{otherwise}
        \end{cases}
    \end{equation}
    where $r(x_{1}, c)$ is the reward score evaluated on the final sample $x_{1}$.
\end{itemize}

Formulating FM as an MDP enables RL-based velocity field refinement. However, deterministic inference ($\pi(a_t \mid s_t)$ as Dirac delta) limits exploration. We address this by converting ODE to SDE (Section 3.2), introducing stochasticity for effective RL training.

\subsubsection{ODE to SDE}

Flow‑GRPO\cite{flowgrpo} converts the deterministic ODE sampler into an equivalent SDE that preserves marginal distributions while introducing GRPO's required stochasticity. The standard FM model uses the deterministic ODE:

\begin{equation}
\begin{array}{ccc}
dx_{t} = v_{t} dt
\end{array}
\end{equation}

which is converted to the reverse‑time SDE:

\begin{equation}
\begin{array}{ccc}
dx_{t} = (v_{t}(x_{t}) - \frac{\sigma_{t}^{2}}{2} \nabla \log p_{t}(x_{t}))dt + \sigma_{t} dw
\end{array}
\end{equation}

According to \cite{flowgrpo}, the above expression can be transformed into:

\begin{equation}
\begin{array}{ccc}
dx_{t} = [v_{t}(x_{t}) + \frac{\sigma_{t}^{2}}{2(1-t)}(x_{t} + tv_{t}(x_{t}))dt ] + \sigma_{t} dw
\end{array}
\end{equation}

The final update rule in \cite{flowgrpo} is:

\begin{equation}
\label{eq:equ1}
\begin{array}{ccc}
x_{t + \Delta t} = x_{t, \mathrm{mean}} + \sigma_{t} \sqrt{\Delta t} \epsilon \\
x_{t, \mathrm{mean}} = x_{t} + [ v_{\theta}(x_{t}, t) + \frac{\sigma_{t}^{2}}{2t} (x_{t} + (1 - t)v_{\theta}(x_{t}, t)) ] \Delta t
\end{array}
\end{equation}

Because our FM uses the opposite time ordering to the original Flow-GRPO and is the same as FlowSE-GRPO, we change Equation \ref{eq:equ1} to:

\begin{equation}
\label{eq:sde}
\begin{aligned}
x_{t, \mathrm{mean}} = x_{t} + \big[& v_{\theta}(x_{t}, t) \\
& + \frac{\sigma_{t}^{2}}{2(1-t)} (-x_{t} + t v_{\theta}(x_{t}, t)) \big] \Delta t .
\end{aligned}
\end{equation}

\noindent where $\epsilon\sim\mathcal{N}(0,\mathbf{I})$ and $\sigma_{t}=a\sqrt{\frac{1-t}{t}}$. The noise level $a$ is a hyperparameter that controls stochasticity in the reverse‑time denoising process.


Inspired by FlowSE-GRPO\cite{wang2026flowse}, MixGRPO\cite{li2025mixgrpo} and Flow-GRPO-Fast\cite{flowgrpo}, we also use window training. We apply SDE sampling only to a subset of early steps $t \in S = [S_{min}, S_{min} + ws]$, where $S_{min}$ denotes the SDE window start and $ws$ denotes the window size. The remaining steps use deterministic ODE updates. Optimization is performed only on these stochastic steps to reduce training burden and accelerate convergence.

\subsubsection{GRPO Loss}

RL learns a policy that maximizes the expected cumulative reward. The policy optimization objective is defined as:

\begin{equation}
\begin{array}{ccc}
\mathrm{max}_{\theta} E_{(s_{0}, a_{0}, ...) \sim \pi_{\theta}} \big[ \sum_{t \in S} (R(s_{t}, a_{t}) \\ - \beta D_{KL}(\pi_{\theta}(\cdot | s_{t}) || \pi_{ref}(\cdot | s_{t})  ) )\big]
\end{array}
\end{equation}


GRPO estimates the advantage $A$ using an intra-group relative formulation, which computes advantages by comparing each sample's reward to the group mean. As shown in Fig \ref{fig:grpo}, given all zero-shot TTS conditions $c$, the FM TTS model $p_{\theta}$ samples a group of $G$ candidate outputs $\{\hat{x}_{T=1}^{i}\}_{i=1}^{G}$ with the mixed ODE–SDE sampler, together with their trajectories $\{(x_{0}^{i},\ldots,x_{T-1}^{i},x_{T=1}^{i})\}_{i=1}^{G}$. The advantage for the $i$‑th estimated audio is then computed from the group‑wise policy rewards.

\begin{equation}
\begin{array}{ccc}
\hat A_{t}^{i} = \frac{ R(\hat x_{1}^{i}, c) - \mathrm{mean}(\{R(\hat x_{1}^{i}, c)\}_{i=1}^{G}) } { \mathrm{std}(\{R(\hat x_{1}^{i}, c)\}_{i=1}^{G}) }
\end{array}
\end{equation}

GRPO updates the policy model by maximizing the following objective:

\begin{equation}
\begin{array}{ccc}
\mathcal{J}_{\mathrm{flow\text{-}grpo}}(\theta) = E_{c \sim C, \{x_{i}\}_{i=1}^{G} \sim \pi_{\theta_{\mathrm{old}}(\cdot | c) } }f(r, \hat A, \theta, \epsilon, \beta), \\

f(r, \hat A, \theta, \epsilon, \beta) = \frac{1}{G} \sum_{i=1}^{G} \frac{1}{S} \sum_{t \in S} \big(  \\ 
\mathrm{min}( r_{t}^{i}(\theta)\hat A_{t}^{i}, \mathrm{clip}(r_{t}^{i}(\theta), 1-\epsilon, 1+\epsilon) \hat A_{t}^{i} ) \big) \\
- \beta D_{\mathrm{KL}}(\pi_{\theta} || \pi_{\theta_{\mathrm{ref}}}) )

\end{array}
\end{equation}

\noindent where
$
r_{t}^{i}(\theta)=\frac{p_{\theta}(x_{t+1}^{i}\mid x_{t}^{i},c)}{p_{\theta_{\mathrm{old}}}(x_{t+1}^{i}\mid x_{t}^{i},c)}.
$
The likelihood $p_{\theta}(x_{t+1}^{i}\mid x_{t}^{i},c) = p_{\theta}(x_{t+\Delta t}^{i}\mid x_{t}^{i},c)= \exp (\log p_{\theta}(x_{t+\Delta t}^{i} \mid x_{t}^{i}, c) )$ is the Gaussian density, which is defined as:

\begin{equation}
\begin{array}{ccc}
\log p_{\theta}(x_{t+\Delta t}^{i} \mid x_{t}^{i}, c) = \left( - \frac{\lVert x_{t+\Delta t}^{i} - x_{t, \mathrm{mean}}^{i} \rVert^{2}}{2 \sigma_{t}^{2} \Delta t} \right) \\ - \log(\sigma_{t}  \sqrt{\Delta t}) - \log(\sqrt{2\pi})
\end{array}
\end{equation}


where the mean is defined by the drift term $x_{t, \mathrm{mean}}^{i}$, and $\sigma_{t} \sqrt{\Delta t}$ represents the standard deviation of the added Gaussian noise at each step.


In this framework, the FlowTTS-GRPO maximizes the objective $\mathcal{J}_{\mathrm{flow\text{-}grpo}}(\theta)$ and updates its parameters $\theta$ by adjusting $p_{\theta}(x_{t+\Delta t}^{i} \mid x_{t}^{i}, c)$ such that trajectories resulting in higher rewards $r(x_1, c)$ are assigned higher probabilities, effectively refining the velocity field $\phi_{\theta}$ toward perceptually superior speech synthesis.

\subsection{Reward Function for TTS}

To optimize the FM-based TTS model using GRPO, we design several reward functions that account for speaker identity, linguistic accuracy, and perceptual quality. The components are defined as follows:

\begin{itemize} 

\item Speaker Similarity Reward ($R_{\text{SS}}$): This metric evaluates the timbre consistency between the synthesized speech and the original reference audio. Following CV3\cite{du2025cosyvoice3}, we use ERes2Net\cite{ERes2Net} to extract speaker embeddings from both the generated and reference waveforms. The reward is defined as the cosine similarity between these two embeddings, mapping to the range $[0, 1]$.

\item ASR-based Reward ($R_{\text{asr}}$): To ensure semantic consistency and intelligibility, we utilize an ASR reward. We use off-the-shelf ASR models—Paraformer \cite{paraformer} for Chinese and Whisper-v3 \cite{systran_fwhisper_large_v3} for English—to transcribe the generated audio. The reward is formulated as $1 - \text{CER}$ for Chinese and $1 - \text{WER}$ for English. CER represents character error rate.

\item Perceptual Quality Reward ($R_{\text{mos}}$): Following \cite{du2025cosyvoice3}, we incorporate the P.835 DNSMOS \cite{reddy2022dnsmos835} to estimate the perceived quality of the audio signal. The DNSMOS model predicts three scores based on the P.835 standard: speech quality (SIG), background noise quality (BAK), and overall quality (OVRL). We adopt the P.835 DNSMOS OVRL score as the training reward to jointly optimize for speech naturalness and noise suppression. All generated waveforms are resampled to $16$ kHz before being processed by the DNSMOS model.

\end{itemize}

\subsection{Multi-Objective Reward Optimization}

During the training phase, we observe that optimizing for a single reward often leads to reward hacking, where the model aggressively maximizes the specific proxy reward at the expense of other essential metrics. To mitigate this and balance multiple quality constraints, we investigate two strategies for reward fusion: probabilistic combination and weighted combination.

\subsubsection{Probabilistic combination}

Inspired by PromptRL\cite{wang2026promptrl}, when optimizing multiple objectives, we probabilistically assign each prompt to one reward function. For a prompt that produces a set of G samples, only that reward is used to compute the rewards while all other rewards are set to zero. This ensures the within-group advantage is governed by a single reward, removing the need to account for scale differences between rewards and avoiding reward interference. Furthermore, we propose using weighted probabilities for prompt-to-reward assignment so that prompts are assigned according to different weights; this enables more targeted optimization of particular downstream metrics and allows different emphasis across objectives.

\begin{figure}[!t]
	\centering
\includegraphics[width=1.0\linewidth]{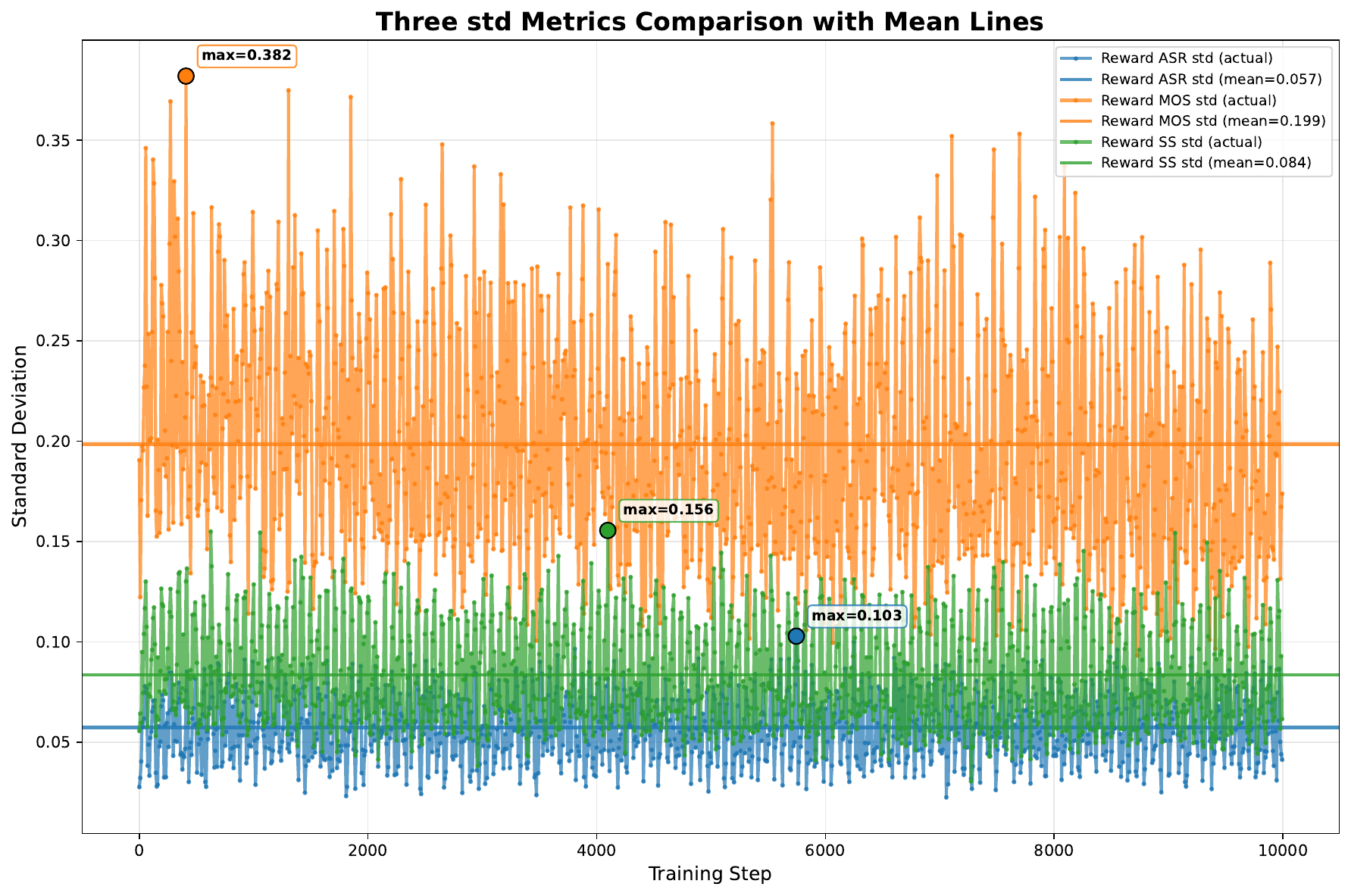}
	\caption{
        The standard deviation of three rewards in batch during training.
	}
	\label{fig:std_compare}
\end{figure}

\subsubsection{Weighted combination}

As shown in Fig.\ref{fig:std_compare}, the standard deviations (std) of different rewards are not equal. If multiple rewards are combined directly by a weighted sum, their contributions to the resulting within-group advantage will not follow the intended weights. Therefore, following FlowSE-GRPO\cite{wang2026flowse}, we normalize each reward by the standard deviation computed over all samples in the current batch, scaling each reward’s std to 1. This allows meaningful design of reward weights, and we use the weighted sum of the normalized rewards as the multi-objective reward:

\begin{equation}
\begin{array}{rcl}
R = \lambda_{1} \frac{R_{\text{SS}}}{\mathrm{std}(R_{\text{SS}})} + \lambda_{2} \frac{R_{\text{ASR}}}{\mathrm{std}(R_{\text{ASR}})} + \lambda_{3} \frac{R_{\text{MOS}}}{\mathrm{std}(R_{\text{MOS}})}
\end{array} 
\end{equation}

\subsection{Robust Training via Hard Case Synthesis}
\label{sec:hardcase}

To enhance the performance of RL on edge cases and improve the model's robustness against complex linguistic patterns, we incorporate a difficulty-aware training strategy. Following the observations in Seed-TTS-Eval \cite{anastassiou2024seedtts}, particularly the HardZh subset which contains challenging patterns such as word repetitions and tongue twisters, we observe that models often struggle with stability and prosodic consistency in these scenarios. To bridge this gap during the RL fine-tuning stage, we curate a specialized "Hard Case" training set by applying heuristic-based text augmentation to the WenetSpeech4TTS dataset. 

For each text sentence, we randomly apply one of the following augmentation strategies to simulate common failure modes: 
\begin{itemize} 
\item \textbf{Local Word Repetition (LWR)}: A single non-punctuation word is randomly selected and repeated $3$ to $5$ times. 
\item \textbf{Sparse Multi-word Repetition (SMR)}: Two or three non-punctuation words are selected, each being repeated $2$ to $3$ times. 
\item \textbf{Global Sentence Repetition (GSR)}: The entire sentence is repeated multiple times to test the model's long-form generation stability. 
\end{itemize} 

Table~\ref{tab:hardcase_examples} shows examples of these augmentation strategies. These augmented texts are then randomly paired with original Chinese prompt waveforms and their corresponding prompt transcriptions. By exposing the agent to these synthesized difficult samples during the FlowTTS-GRPO process, the model learns to maintain high synthesis quality and alignment accuracy even under extreme textual conditions.

\begin{table}[!t]
\centering
\caption{Examples of hard case text augmentation strategies.}
\begin{tabular}{ll}
\toprule
\textbf{Strategy} & \textbf{Example} \\
\midrule
Original Text & 
\begin{CJK*}{UTF8}{gbsn}克隆音色与风格\end{CJK*} \\
\midrule
LWR & \begin{CJK*}{UTF8}{gbsn}克隆音色与与与与风格\end{CJK*}  \\
\midrule
SMR & \begin{CJK*}{UTF8}{gbsn}克隆克隆音色音色音色与风格\end{CJK*}  \\
\midrule
GSR & \begin{CJK*}{UTF8}{gbsn}克隆音色与风格，克隆音色与风格\end{CJK*}  \\
\bottomrule
\end{tabular}
\label{tab:hardcase_examples}
\end{table}

\begin{table*}[!t]
    \centering
    \caption{Zero-shot TTS performance comparison on Seed-TTS-Eval. SS1: WavLM-based speaker similarity; SS2: ERes2Net-based speaker similarity; P808: P.808 DNSMOS; P835: P.835 DNSMOS (proxy reward); w/o: without; w.: with; Step: RL training steps.
    }
    \resizebox{1.0\textwidth}{!}{
    {
    \begin{tabular}{lccccccccccccccccc}
        \toprule
			\multirow{2}{*}{\textbf{Model}}  & \multirow{2}{*}{\textbf{RL}}  & \multirow{2}{*}{\textbf{Step}} & \multicolumn{5}{c}{\textbf{\emph{test-zh}}} & \multicolumn{5}{c}{\textbf{\emph{test-en}}} & \multicolumn{5}{c}{\textbf{\emph{test-hard}}}\\
            
			\cmidrule(r){4-8} \cmidrule(r){9-13} \cmidrule(r){14-18}
			& & & \textbf{CER~$\downarrow$} & \multicolumn{1}{c}{\textbf{SS1~$\uparrow$}} & \multicolumn{1}{c}{\textbf{SS2~$\uparrow$}} & \multicolumn{1}{c}{\textbf{P808~$\uparrow$}} & \multicolumn{1}{c}{\textbf{P835~$\uparrow$}} & \textbf{WER~$\downarrow$} & \multicolumn{1}{c}{\textbf{SS1~$\uparrow$}} & \multicolumn{1}{c}{\textbf{SS2~$\uparrow$}} & \multicolumn{1}{c}{\textbf{P808~$\uparrow$}} & \multicolumn{1}{c}{\textbf{P835~$\uparrow$}} & \textbf{CER~$\downarrow$} & \multicolumn{1}{c}{\textbf{SS1~$\uparrow$}} & \multicolumn{1}{c}{\textbf{SS2~$\uparrow$}} & \multicolumn{1}{c}{\textbf{P808~$\uparrow$}} & \multicolumn{1}{c}{\textbf{P835~$\uparrow$}} \\
			\midrule
			\textbf{Human}& & & 1.26 & 0.755 & 0.775  &  - & - & 2.14 & 0.734 & 0.742   &  - & - & - & -  &  - & - \\
            \midrule
            \multicolumn{18}{c}{\textbf{TTS Models w/o. RL}} \\
             \midrule
            F5-TTS~\cite{chen2025f5tts} & &  & 1.56 & 0.741 & 0.794  &  - & - & 1.83 & 0.647 & 0.742  &  - & - & 8.67 & 0.713 &0.762  &  - & - \\
            MaskGCT~\cite{wang2024maskgct} & &  & 2.27 & 0.774 & 0.752  &  - & - & 2.62 &  0.714 & 0.730  &  - & - & 10.27 & 0.748 & 0.720 &  - & -  \\
            Seed-TTS~\cite{anastassiou2024seedtts} & &  & 1.12 & 0.796 & - & - &  - & 2.25 & \textbf{0.762} & - &  - & -  & 7.59 & 0.776 & - &  - & -\\
            CosyVoice~\cite{du2024cosyvoice1} & &    & 3.63 & 0.723 & 0.775 & - &  - & 4.29 & 0.609 & 0.699 & - &  - & 11.75  & 0.709 & 0.755 & - &  -\\
            CosyVoice 2.0~\cite{du2024cosyvoice2} & &    & 1.45 & 0.748 & 0.806 & - &  - & 2.57 & 0.652 & 0.736 & - &  - & 6.83 & 0.724 & 0.776 & - &  - \\

           \midrule
            \multicolumn{18}{c}{\textbf{TTS Models w. RL on LLM}} \\
             \midrule

        \multirow{2}{*}{Llasa-1B~\cite{ye2025llasa}} & \ding{55} &  & 7.73 & 0.636 & - & - &  - & 4.95 &  0.578 &  - & - &  - &  - & - &  - & - &  - \\
    & LM-GRPO &  & 1.30 & 0.669 & -  & - &  - & 2.17 &  0.580 &  - & -  & - &  - & - &  -  & - &  - \\

        \multirow{2}{*}{CosyVoice 2.0~\cite{du2024cosyvoice2}} & \ding{55} &  & 1.41 & 0.753 & -  & - &  - & 2.46 & 0.655 &  -  & - &  - &  - & - &  -  & - &  - \\
    & LM-GRPO &  & 1.07 & 0.753 & -  & - &  - & 2.30 & 0.659 &  -  & - &  - & - & - &  -  & - &  - \\

        \multirow{2}{*}{CosyVoice 3.0-0.5B~\cite{du2025cosyvoice3}} & \ding{55} &  & 1.16 & 0.780 & 0.840 & - &  - & 2.02 &  0.718&  0.790 & - &  - &  6.08 & 0.758 &  0.815& - &  - \\
    & LM-DiffRO &  & 0.75 & 0.774 & 0.836& - &  -  &  1.76 & 0.695 & 0.783 & - &  - & \textbf{5.09} & 0.750 &  0.809 & - &  -\\
        \multirow{2}{*}{CosyVoice 3.0-1.5B~\cite{du2025cosyvoice3}} & \ding{55} &  & 1.12 & 0.781  & 0.837 & - &  - & 2.21  & 0.720  & 0.789 & - &  - & 5.83  & 0.758  & 0.816 & - &  - \\
    & LM-DiffRO &  & 0.71 & 0.775 & 0.836 & - &  - & \textbf{1.45} & 0.695 & 0.784 & - &  - & 5.66&  0.750 & 0.810 & - &  - \\

           \midrule
            \multicolumn{18}{c}{\textbf{TTS Models w. RL on FM}} \\
             \midrule
        \multirow{2}{*}{F5R-TTS~\cite{sun2025f5r}} & \ding{55} &  & 1.65 & 0.726 & -  & - &  - & - & - &  -  & - &  - & 9.87 & 0.702 &  -  & - &  -\\
    & FM-GRPO &  & 1.37 & 0.754 & -  & - &  - & - &  - & -  & - &  - & 8.79 & 0.718 & -  & - &  - \\
            \midrule
            \multicolumn{18}{c}{\textbf{Ours FlowTTS-GRPO Models}} \\
             \midrule
        \multirow{2}{*}{F5-TTS} & \ding{55} & 0 & 1.81 & 0.760 &  0.796  & 3.762 & 3.313 & 1.88 & 0.677 & 0.753  & 3.794 & 3.154 & 9.00 & 0.730 & 0.759 & 3.765 & 3.304 \\
    & FM-GRPO & 1289 & 1.55 & 0.777 & 0.827 & 3.948 & 3.514 & 1.73	& 0.705 & 0.790 & 3.915 & 3.408 & 7.86 & 0.741 & 0.791 & 3.973 & 3.562 \\
                 
                \cmidrule(lr){1-18}
                 
        \multirow{2}{*}{CosyVoice 3.0-0.5B-2512} & \ding{55} & 0 & 1.20 & 0.777 & 0.830   & 3.889 &  3.353 &	2.42 &	0.701 &	0.770 & 3.910 &  3.226 & 7.32 &	0.757 & 0.808  & 3.875 &  3.393  \\
    & FM-GRPO & 9545 & 1.26 & \textbf{0.804} & \textbf{0.859} & \textbf{3.987} & \textbf{3.536} & 2.49	& 0.743 & \textbf{0.818} & \textbf{3.956} & 	\textbf{3.460} & 7.08 & \textbf{0.792} & \textbf{0.844} & \textbf{3.976} & \textbf{3.559} \\
                 
                 \cmidrule(lr){1-18}
                 
        \multirow{2}{*}{\makecell[l]{CosyVoice 3.0-0.5B-2512 \\ \ \ + LM w. RL}} & \ding{55} &  0 & 0.87 &  0.776 & 0.831  & 3.878 &  3.347  & 1.70 & 0.693 & 0.770   & 3.903 &  3.231 & 6.01 & 0.756 & 0.803  & 3.871 &  3.390 \\
    & FM-GRPO &  9545 & \textbf{0.85} & 0.803 & 0.858 & 3.979	 & 3.528 & 1.83	& 0.737 & 0.817	& 3.954 & 3.457 & 5.89 & 0.790 & 0.841 & 3.971 & 3.544 \\

        \bottomrule
    \end{tabular}}}
    \label{tab:main}
\end{table*}

\begin{figure*}[!t]
	\centering
\includegraphics[width=0.95\linewidth]{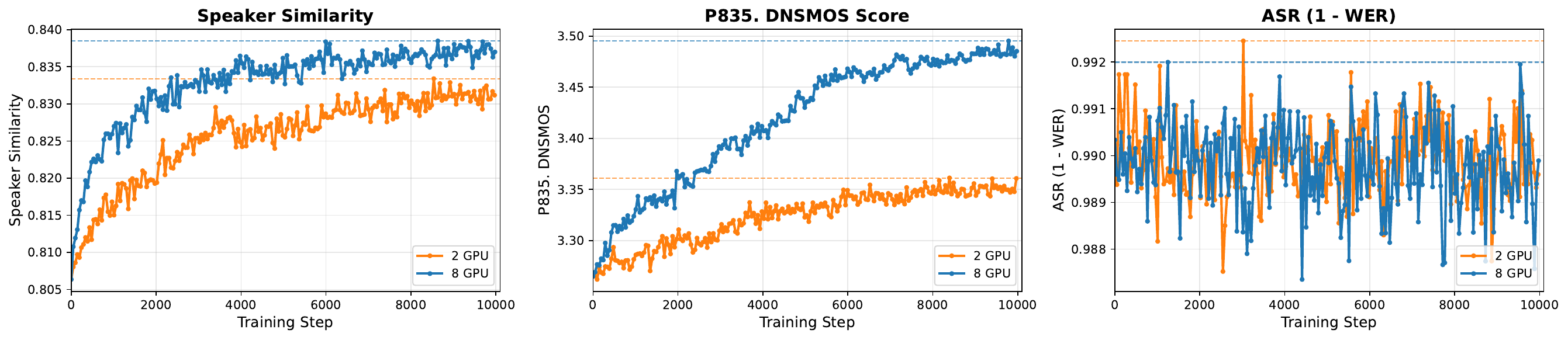}
	\caption{
        Proxy reward curves for CV3 on the dev-easy set during RL training with different numbers of GPUs.
	}
	\label{fig:scale_gpu_pic}
\end{figure*}

\section{Experimental Setup}

\subsection{Training Dataset}

We use WenetSpeech4TTS\cite{ma2024wenetspeech4tts} Premium (Chinese) and LibriTTS-960\cite{libritts} (English) as training sets. Audio files serve as prompt waveforms with transcripts as prompt text. We randomly shuffle the original text corpus to produce target texts for voice cloning. We construct 20k samples each for Chinese and English (40k total, named train-easy-40k). Additionally, following Section \ref{sec:hardcase}, we generate 20k extra challenging Chinese training samples (train-hard-20k).

\subsection{Validation Sets}

We construct two validation sets: dev-easy (200 utterances, randomly selected from Seed-TTS-Eval\cite{anastassiou2024seedtts} Chinese and English test sets) and dev-hard (200 samples from Seed-TTS-Eval hard-zh test set with repeated-text cases). dev-easy monitors metric trends, while dev-hard specifically tracks CER changes on challenging examples.

\subsection{Training and Model Configuration}

We fine-tune CosyVoice 3.0\cite{du2025cosyvoice3} (CV3-2512 version) and F5-TTS\cite{chen2025f5tts} (Base\_v1) with 8 GPUs. For CV3, we pre-decode the training set with the 0.5B LM front-end (not RL-trained) to extract prompt and generated tokens for FM training. F5-TTS is fine-tuned directly.

For CV3, we use Low-Rank Adaptation (LoRA) \cite{hu2022lora} with rank 32 and lora\_alpha 64 (10.09M trainable parameters, 2.78\% of FM model). Learning rate is 1e-4 with linear decay to 0 at 10k steps. Each iteration samples 16 prompt waveforms (repeated 4 times), generates 8 samples per prompt (512 total). We discard groups with std=0, reassemble into batch size 16, and perform 8 parameter updates per iteration. Training uses 2-step window (denoising step in [5, 8], Smin in [1, 3]) without CFG; inference uses 10 denoising steps. Unless otherwise noted, main training uses train-easy-40k and dev-easy. 

For F5-TTS, we use LoRA with rank 32 and lora\_alpha 64 (10.09M trainable parameters, 2.80\% of FM model). Learning rate is 5e-5 with linear decay to 0 at 10k steps. Each iteration samples 6 prompt waveforms (repeated 4 times), generates 10 samples per prompt (240 total). We discard groups with std=0, reassemble into batch size 6, and perform 8 parameter updates per iteration. Training uses 2-step window (denoising step in [8, 16], Smin in [1, 3]) without CFG; inference uses 16 denoising steps. Unless otherwise noted, main training uses train-easy-40k + train-hard-20k and dev-hard.

\subsection{Evaluation Sets}

We evaluate on Seed‑TTS‑Eval\cite{anastassiou2024seedtts} (Chinese test-zh: 2,020 samples; English test-en: 1,088 samples; challenging test-hard: 400 samples) and CV3‑Eval\cite{du2025cosyvoice3}. CV3‑Eval uses the Multilingual Voice Cloning subset, which contains nine languages with 500 samples each: Chinese (zh), English (en), Japanese (ja), Korean (ko), German (de), French (fr), Russian (ru), Italian (it), and Spanish (es). We also include two additional hard‑case test sets from CV3‑Eval for Chinese and English.

\begin{table*}[t]\centering
\small
\setlength{\tabcolsep}{0.8mm}{
  \centering
  \footnotesize
  \caption{
        \label{tab:cv3_eval}
  {Zero-shot TTS performance comparison between F5-TTS, CosyVoice 3.0 with FlowTTS-GRPO RL on the CV3-Eval Multilingual Voice Cloning subset. Step represents RL training steps.
}}
  \begin{tabular}{cccccccccccccc}
  \toprule
  \textbf{Metric}  & \textbf{Models}  & Step & zh & en & hard-zh & hard-en & ja & ko & de & es & fr & it & ru \\
  
  \midrule


  \multirow{6}*{CER} & \multirow{2}*{F5-TTS} & 0 & 7.66 & 11.10 & 22.11 & 36.23 & - & - & - & - & - & - & - \\

  & &  \cellcolor{lightgray}1289 & \cellcolor{lightgray}5.44 & \cellcolor{lightgray}7.07 & \cellcolor{lightgray}17.46 & \cellcolor{lightgray}25.63 & \cellcolor{lightgray}- & \cellcolor{lightgray}- & \cellcolor{lightgray}- &\cellcolor{lightgray}- & \cellcolor{lightgray}- & \cellcolor{lightgray}- & \cellcolor{lightgray}- \\

  & \multirow{2}*{CV3} &  0 & 3.72 & 5.09 & 9.04 & 8.11 & 6.54 & 5.92 & 6.48 & 3.58 & 10.66 & 5.43 & 7.84 \\

  & &  \cellcolor{lightgray}9545 & \cellcolor{lightgray}3.83 & \cellcolor{lightgray}5.09 & \cellcolor{lightgray}9.38 &	\cellcolor{lightgray}7.93 &	\cellcolor{lightgray}6.72 &	\cellcolor{lightgray}5.85 &	\cellcolor{lightgray}6.79 &	\cellcolor{lightgray}3.64 &	\cellcolor{lightgray}11.69 &	\cellcolor{lightgray}5.74 &	\cellcolor{lightgray}8.29  \\

  & \multirow{2}*{CV3-LM-RL} &  0 & 3.26 & 4.10 & 8.39 & 7.30 &	5.77 &	6.28 & 4.72 & 3.15 & 8.63 & 3.36 & 4.39 \\

  & &  \cellcolor{lightgray}9545 & \cellcolor{lightgray}3.28	& \cellcolor{lightgray}4.31 & \cellcolor{lightgray}8.26 & \cellcolor{lightgray}7.29 & \cellcolor{lightgray}5.91 & \cellcolor{lightgray}6.10 & \cellcolor{lightgray}4.69 & \cellcolor{lightgray}3.40 & \cellcolor{lightgray}8.52 & \cellcolor{lightgray}3.75 &\cellcolor{lightgray}4.67 \\

  \midrule

  \multirow{6}*{SS1} & \multirow{2}*{F5-TTS} & 0 & 0.729	&0.605	&0.688	&0.559 &	-	&-	&-	&-&	-	&-	&- \\

  & &  \cellcolor{lightgray}1289 & \cellcolor{lightgray}0.760 &	\cellcolor{lightgray}0.665	&\cellcolor{lightgray}0.718	&\cellcolor{lightgray}0.627&	\cellcolor{lightgray}-&	\cellcolor{lightgray}-	&\cellcolor{lightgray}-	& \cellcolor{lightgray}-&	\cellcolor{lightgray}-	&\cellcolor{lightgray}-	&\cellcolor{lightgray}- \\

  & \multirow{2}*{CV3} &  0 &  0.777	&0.681	&0.765	&0.687	&0.748	&0.761	&0.735	&0.752&	0.718	&0.732	&0.745 \\

  & &  \cellcolor{lightgray}9545 & \cellcolor{lightgray}0.803&	\cellcolor{lightgray}0.725	&\cellcolor{lightgray}0.794	&\cellcolor{lightgray}0.736&	\cellcolor{lightgray}0.783	&\cellcolor{lightgray}0.791	&\cellcolor{lightgray}0.776	&\cellcolor{lightgray}0.787	&\cellcolor{lightgray}0.764	&\cellcolor{lightgray}0.773&	\cellcolor{lightgray}0.784  \\

  & \multirow{2}*{CV3-LM-RL} &  0 & 0.773&	0.658	&0.761	&0.666	&0.740&	0.738&	0.726	&0.741&	0.707	&0.725	&0.743 \\

  & &  \cellcolor{lightgray}9545 & \cellcolor{lightgray}0.800	&\cellcolor{lightgray}0.708	&\cellcolor{lightgray}0.787	&\cellcolor{lightgray}0.719	&\cellcolor{lightgray}0.774	&\cellcolor{lightgray}0.774&	\cellcolor{lightgray}0.770	&\cellcolor{lightgray}0.781	&\cellcolor{lightgray}0.756	&\cellcolor{lightgray}0.769	&\cellcolor{lightgray}0.783 \\

  \midrule

  \multirow{6}*{SS2} & \multirow{2}*{F5-TTS} & 0 & 0.729 & 0.655 & 0.688 & 0.618 &	-	&-	&-	&-&	-	&-	&- \\

  & &  \cellcolor{lightgray}1289 & \cellcolor{lightgray}0.799 &	\cellcolor{lightgray}0.755	& \cellcolor{lightgray}0.761 & \cellcolor{lightgray}0.727 & \cellcolor{lightgray}- &	\cellcolor{lightgray}-	&\cellcolor{lightgray}-	& \cellcolor{lightgray}-&	\cellcolor{lightgray}-	&\cellcolor{lightgray}-	&\cellcolor{lightgray}- \\

  & \multirow{2}*{CV3} &  0 &  0.804 &	0.745 &	0.780 &	0.747 & 0.777 &	0.800 &	0.785 &	0.796 &	0.772 &0.772 & 0.775 \\

  & &  \cellcolor{lightgray}9545 & \cellcolor{lightgray}0.836	&\cellcolor{lightgray}0.798	&\cellcolor{lightgray}0.817	&\cellcolor{lightgray}0.804	&\cellcolor{lightgray}0.816	&\cellcolor{lightgray}0.834&	\cellcolor{lightgray}0.827&	\cellcolor{lightgray}0.838	& \cellcolor{lightgray}0.816&	\cellcolor{lightgray}0.820	&\cellcolor{lightgray}0.820  \\

  & \multirow{2}*{CV3-LM-RL} &  0 & 0.798	& 0.737 & 0.776 & 0.737 &	0.773 & 0.777 &	0.780 &	0.787&	0.765&	0.768&	0.773 \\

  & &  \cellcolor{lightgray}9545 & \cellcolor{lightgray}0.832	&\cellcolor{lightgray}0.791	&\cellcolor{lightgray}0.814	& \cellcolor{lightgray}0.797	&\cellcolor{lightgray}0.813	&\cellcolor{lightgray}0.818	&\cellcolor{lightgray}0.823	&\cellcolor{lightgray}0.830	&\cellcolor{lightgray}0.813	&\cellcolor{lightgray}0.819	& \cellcolor{lightgray}0.819 \\

  \midrule

  \multirow{6}*{P808} & \multirow{2}*{F5-TTS} & 0 & 3.358 & 3.300 & 3.224 & 3.220 & - & - & - & - & - &- & - \\

  & &  \cellcolor{lightgray}1289 & \cellcolor{lightgray}3.758 & \cellcolor{lightgray}3.716	& \cellcolor{lightgray}3.662 & \cellcolor{lightgray}3.726 & \cellcolor{lightgray}- & \cellcolor{lightgray}-	&\cellcolor{lightgray}-	&\cellcolor{lightgray}-&	\cellcolor{lightgray}-	&\cellcolor{lightgray}-	&\cellcolor{lightgray}- \\

  & \multirow{2}*{CV3} &  0 &  3.855 &	3.839 & 3.793 & 3.958 &	3.825 & 3.854 &	3.821 & 3.830 & 3.802 & 3.779 & 3.829 \\
 
  & &  \cellcolor{lightgray}9545 & \cellcolor{lightgray}3.923 & \cellcolor{lightgray}3.897 & \cellcolor{lightgray}3.881 & \cellcolor{lightgray}4.007 & \cellcolor{lightgray}3.870 & \cellcolor{lightgray}3.908 & \cellcolor{lightgray}3.884 & \cellcolor{lightgray}3.882 & \cellcolor{lightgray}3.849 & \cellcolor{lightgray}3.829 & \cellcolor{lightgray}3.885  \\

  & \multirow{2}*{CV3-LM-RL} &  0 & 3.864 & 3.835 & 3.826 & 3.966 & 3.799 & 3.790 & 3.806 & 3.827 & 3.786 & 3.758 & 3.831 \\

  & &  \cellcolor{lightgray}9545 & \cellcolor{lightgray}3.931 & \cellcolor{lightgray}3.889 & \cellcolor{lightgray}3.900 & \cellcolor{lightgray}3.997 & \cellcolor{lightgray}3.856 & \cellcolor{lightgray}3.846 & \cellcolor{lightgray}3.880 & \cellcolor{lightgray}3.875 &	\cellcolor{lightgray}3.844 &	\cellcolor{lightgray}3.812 & \cellcolor{lightgray}3.890 \\

  \bottomrule

\end{tabular}
}

\end{table*}

\section{Results and Discussion}

\subsection{Evaluation Metrics}

Following CV3\cite{du2025cosyvoice3}, we evaluate the effect of FlowTTS-GRPO fine-tuning using three objective metrics:

\begin{itemize}
    \item Content consistency (CER/WER): measures the intelligibility of synthesized speech. We report Character Error Rate (CER) for Chinese and other non-English languages, and Word Error Rate (WER) for English. For Chinese we use Paraformer\cite{paraformer} and for other languages we use Whisper-large-v3\cite{systran_fwhisper_large_v3}.
    \item Speaker similarity (SS): measures how well the synthesized audio matches the reference by computing the cosine similarity between their speaker embeddings. We report two speaker similarity metrics: SS1 (WavLM-based)\cite{chen2022large_wavlmspk} and SS2 (ERes2Net-based)\cite{ERes2Net}.
    \item Audio quality: following \cite{du2025cosyvoice3}, we score generated speech using P.808 DNSMOS (P808)\cite{reddy2021dnsmos808} and P.835 DNSMOS (P835)\cite{reddy2022dnsmos835}. The DNSMOS scores correlate highly with human auditory perception; we use P835 as a proxy reward during training.
\end{itemize}

Beyond these objective metrics, we also conduct subjective A/B preference tests in Sec. \ref{sec:subject_eval} to assess whether reward improvements translate to human-perceived naturalness and timbre similarity.

\subsection{Main Results}

\subsubsection{Main Results of CV3 on Seed-TTS-Eval}
\label{sec:main_cv3_seedtts}

The blue curve in Fig. \ref{fig:scale_gpu_pic} shows validation trends for CV3-FM across speaker similarity (ERes2Net-based, SS2), perceptual quality (P835), and intelligibility (ASR reward) under weighted combination multi-objective RL. Both SS2 and P835 improve steadily, demonstrating the effectiveness of GRPO fine-tuning for FM-based TTS. However, the ASR reward plateaus near 0.99, likely because (1) the system already achieves high intelligibility, and (2) in the LLM+FM pipeline, intelligibility is primarily constrained by LLM-generated tokens. This aligns with findings that RL on the LLM component is the main driver for WER gains. Based on the validation $R_{\text{ASR}}$, we select the 9545-step checkpoint for final evaluation. As shown in Tab. \ref{tab:main}, FM-RL (FM-GRPO) improves SS2 from 0.830 to 0.859 and SS1 from 0.777 to 0.804 on Seed-TTS-Eval-zh compared to the baseline CosyVoice 3.0-0.5B-2512 (CV3), with similar gains in English and hard-zh. Notably, our model’s SS1 on Seed-TTS-Eval-zh surpasses the closed-source Seed-TTS for the first time, achieving state-of-the-art (SOTA) performance.

While improvement in SS2 is expected because it serves as a proxy reward model, SS1, which is not used for optimization, also improves. This indicates that RL fine-tuning does not overfit or "hack" to the reward model and that the speaker similarity gains are robust.

CV3 also provides an LM-with-RL variant (CV3-0.5B-2512 + LM w.RL, CV3-LM-RL). Although FM fine-tuning uses tokens from the LM without RL, the RL-tuned FM generalizes well to CV3-LM-RL, improving both speaker similarity and P.835 DNSMOS. This suggests a functional decoupling in LLM+FM systems: RL for intelligibility is most effective on the LM, while RL for audio details (e.g., timbre similarity and perceptual quality) is better suited for the FM.

As shown in Fig. \ref{fig:scale_gpu_pic}, increasing the GPU count increases the number of samples seen per update and accelerates reward growth. This suggests that scaling the data throughput per update enhances RL training effectiveness.

\subsubsection{Main Results of F5-TTS}
Fig. \ref{fig:hardzh_pic} (blue curve) shows validation improvements for F5-TTS in SS2, P.835, and ASR reward. During training, SS2 and P.835 rise significantly, while the ASR metric improves more gradually. We evaluate the 1289-step checkpoint of F5-TTS (F5-TTS required substantially more training time than CV3-FM, and resource limits constrained our runs). As shown in Tab. \ref{tab:main}, FM-RL enhances both in-domain and out-of-domain speaker similarity (SS2, SS1) and perceptual quality (P835, P808). Since F5-TTS synthesizes directly from text, unlike the LLM constrained CV3-FM, it shows clear gains in intelligibility, further validating our RL approach.

\subsection{The Results on CV3-Eval}

Tab. \ref{tab:cv3_eval} shows that FM-RL (applied to CV3-LM and CV3-LM-RL) improves speaker similarity and audio quality in languages beyond Chinese and English. Despite training only on Chinese and English, these gains demonstrate strong cross-lingual generalization, which we plan to extend with multilingual data in future work. Furthermore, F5-TTS after FlowTTS-GRPO training shows reduced CER and improved intelligibility and similarity, further validating the broad effectiveness of our method.

\subsection{Ablation Study}

All ablation study results on CV3 use 2 GPUs for training.

\begin{figure*}[!t]
	\centering
\includegraphics[width=0.95\linewidth]{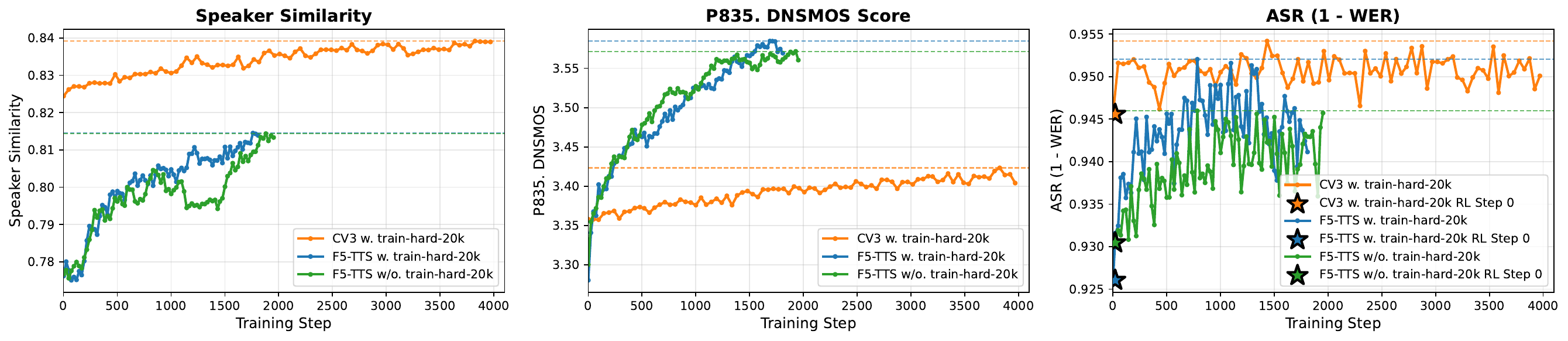}
	\caption{
        Proxy reward curves for CV3 and F5-TTS on the dev-hard set during RL training with Robust Training via Hard Cases. train-hard-20k represents the hard case training set.
	}
	\label{fig:hardzh_pic}
\end{figure*}

\begin{figure*}[!t]
	\centering
\includegraphics[width=0.95\linewidth]{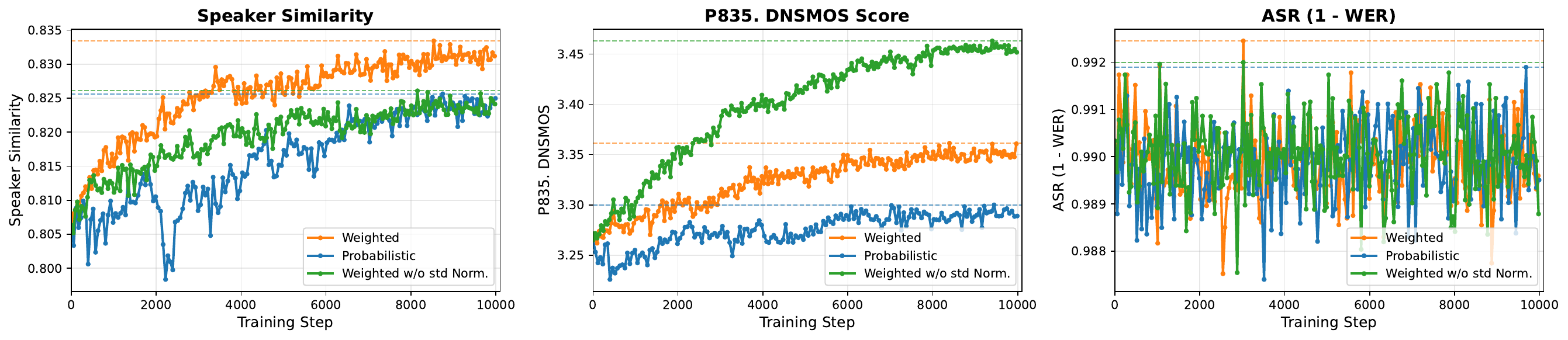}
	\caption{
        Proxy reward curves for CV3 on the dev-easy set during RL training with different multi-objective reward combination strategies.
	}
	\label{fig:multi_reward_methods}
\end{figure*}

\begin{figure*}[!t]
	\centering
\includegraphics[width=0.95\linewidth]{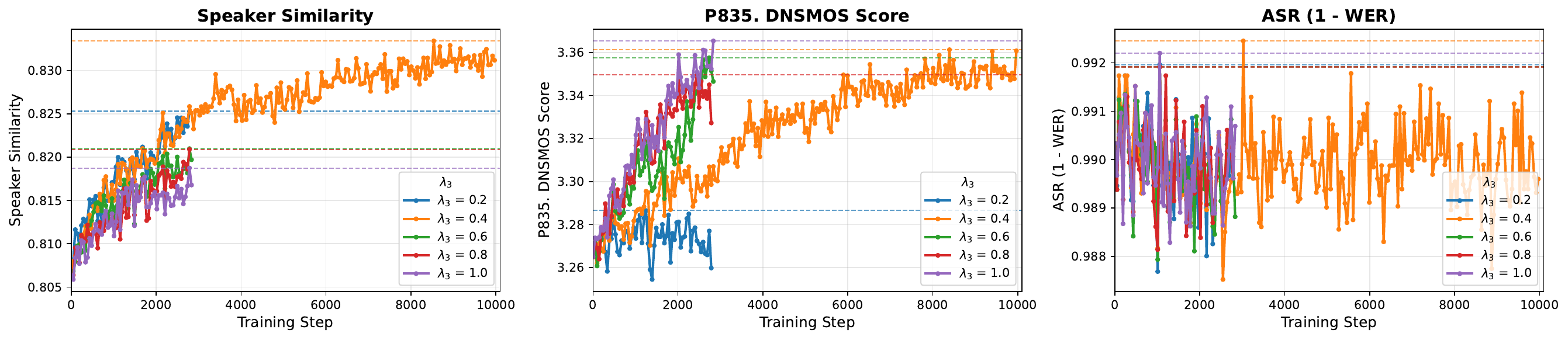}
	\caption{
        Proxy reward curves for CV3 on the dev-easy set during RL training with different DNSMOS weights.
	}
	\label{fig:dnsmos_weight}
\end{figure*}

\begin{figure}[!t]
	\centering
\includegraphics[width=0.95\linewidth]{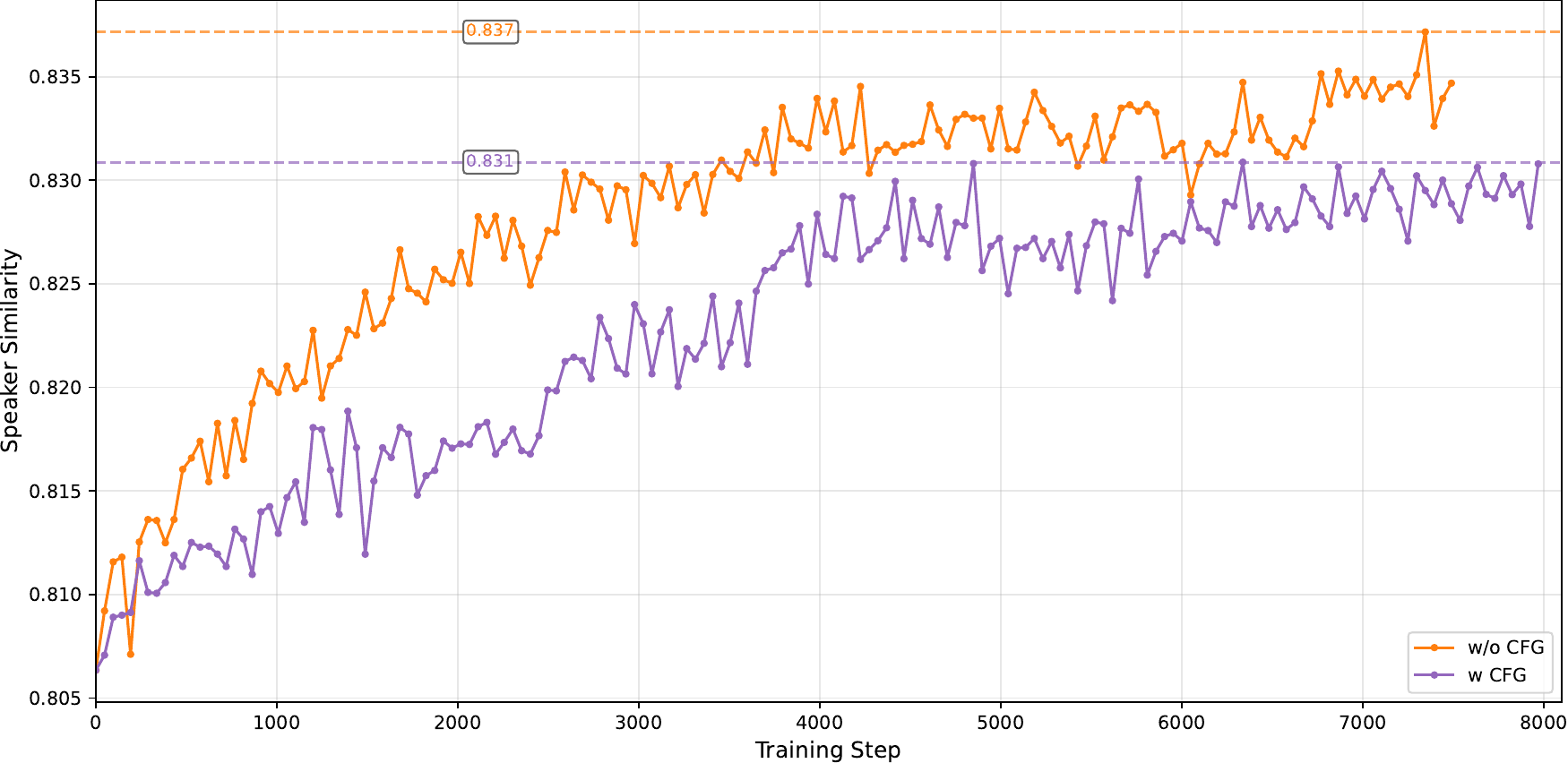}
	\caption{
        Effect of omitting CFG during RL training on the dev-easy set.
	}
	\label{fig:cfg_pic}
\end{figure}

\begin{figure}[!t]
	\centering
\includegraphics[width=0.95\linewidth]{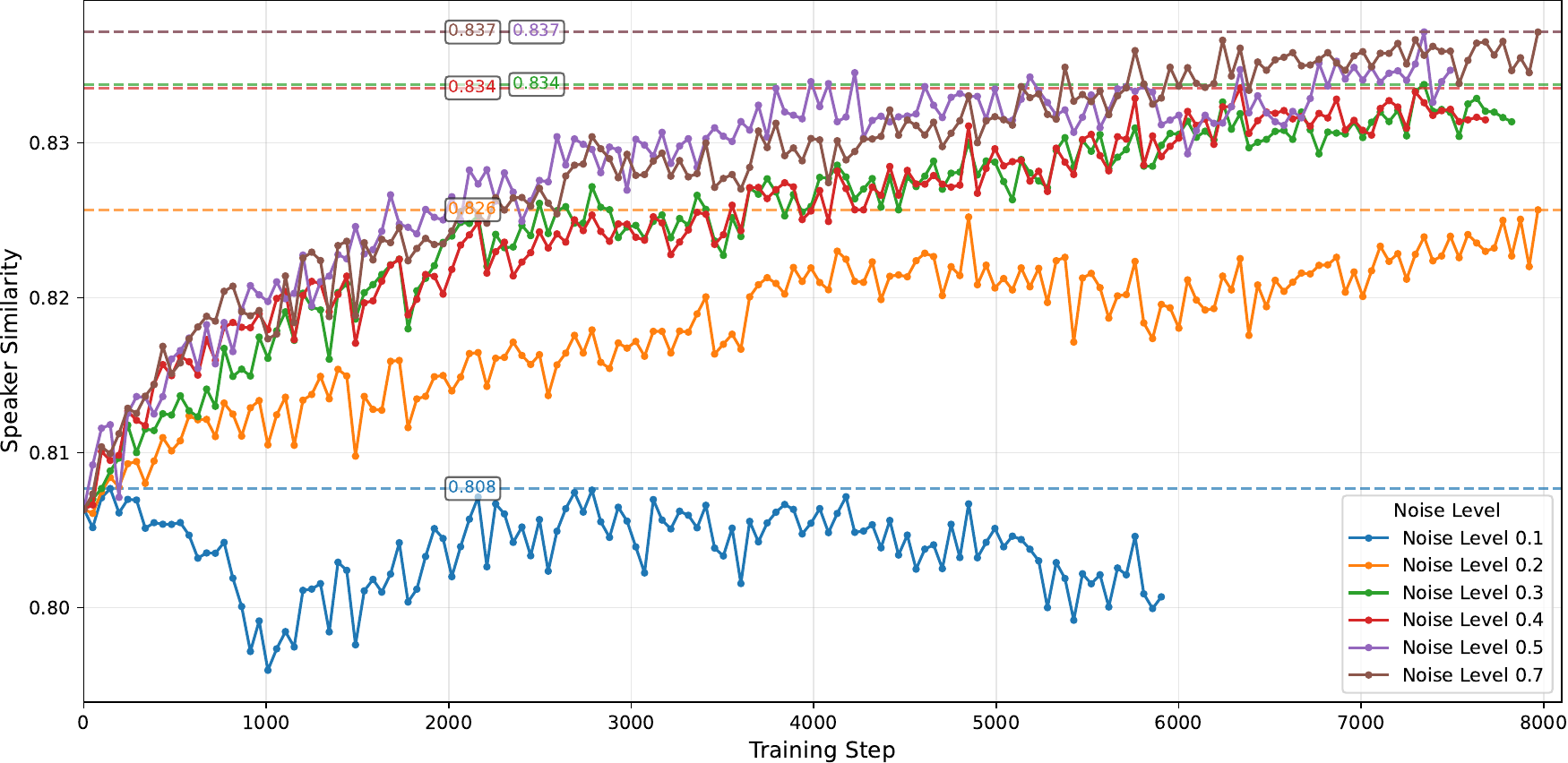}
	\caption{
        Effect of different noise levels during RL training on the dev-easy set.
	}
	\label{fig:noise_level_pic}
\end{figure}

\subsubsection{Robust Training via Hard Case}

Fig. \ref{fig:hardzh_pic} shows proxy-reward trajectories on the dev-hard set for CV3 and F5-TTS variants. For CV3, even adding the train-hard-20k set yields minimal intelligibility gains, suggesting that semantic information is primarily LM-driven. In contrast, F5-TTS synthesizes directly from text as an FM-only architecture and achieves ASR improvements with the train-easy-40k set. This indicates that the FM can independently adapt distribution modeling to enhance intelligibility. Additionally, using train-hard-20k samples accelerates ASR reward growth, supporting the observation that optimization on hard cases is more efficient. We therefore select both train-easy-40k and train-hard-20k for F5-TTS training.

\subsubsection{Different Multi-Objective Reward Combination Strategies}

Fig. \ref{fig:multi_reward_methods} compares the proxy reward trajectories of the probabilistic (orange) and weighted (blue) schemes. The orange curve, which uses per-prompt probabilistic assignment, shows early oscillations because reward preferences vary across groups. While the eventual reward increase validates the probabilistic approach, the weighted combination method yields faster and more stable growth. We therefore adopt the weighted combination method as our final multi-objective strategy.

We also conduct an ablation study on standard deviation (std) normalization during weighted combination. As shown by the green curve in Fig. \ref{fig:multi_reward_methods}, the DNSMOS improvement rate is significantly higher than the orange curve, while the SS growth rate is lower. This result aligns with the fact that $R_{\text{SS}}$ has a lower std than $R_{\text{MOS}}$, which indicates that rewards with higher variance naturally exert more influence on advantage calculation. Therefore, std normalization allows for more precise control over reward contributions, which confirms the effectiveness of this method.

\subsubsection{Different Weights of DNSMOS}

Fig. \ref{fig:dnsmos_weight} shows that adjusting weights in multi-reward combinations regulates the contribution and improvement rate of each metric. Increasing $\lambda_{3}$ from 0.2 to 1.0 accelerates DNSMOS gains but slows speaker similarity growth, which indicates a conflict between their contributions to the advantage. At $\lambda_{3}=0.2$, DNSMOS oscillates near the baseline as its influence is insufficient to drive updates. Given that speaker similarity is critical for zero-shot TTS, we set $\lambda_{3}=0.4$. The configuration $\lambda_{1}=\lambda_{2}=1.0, \lambda_{3}=0.4$ improves similarity while maintaining DNSMOS quality and stable ASR performance.

\subsubsection{Effect of Omitting CFG During Training}

We conduct an ablation study on classifier-free guidance (CFG). Fig. \ref{fig:cfg_pic} compares proxy reward growth when CFG is applied during training sampling versus when it is omitted. This experiment uses SS as the sole reward and applies CFG during evaluation. Training without CFG produces faster proxy reward growth, which indicates that omitting CFG during sampling increases exploration and accelerates RL convergence. This also indicates that optimizing only the conditional velocity still effectively improves performance under CFG-based inference.

\subsubsection{Effect of Noise Levels}

Fig. \ref{fig:noise_level_pic} shows how different noise\_level values influence the proxy reward growth rate. This experiment uses $R_{SS}$ as the sole reward. A higher noise\_level expands the FM’s exploration range under GRPO, which accelerates reward improvement. At noise\_level=0.1, the reward oscillates without sustained gain, whereas increasing the value to 0.5 reaches the training saturation point. We therefore select noise\_level=0.5 because it balances exploration with growth speed.

  







\begin{figure}[!t]
	\centering
\includegraphics[width=1.0\linewidth]{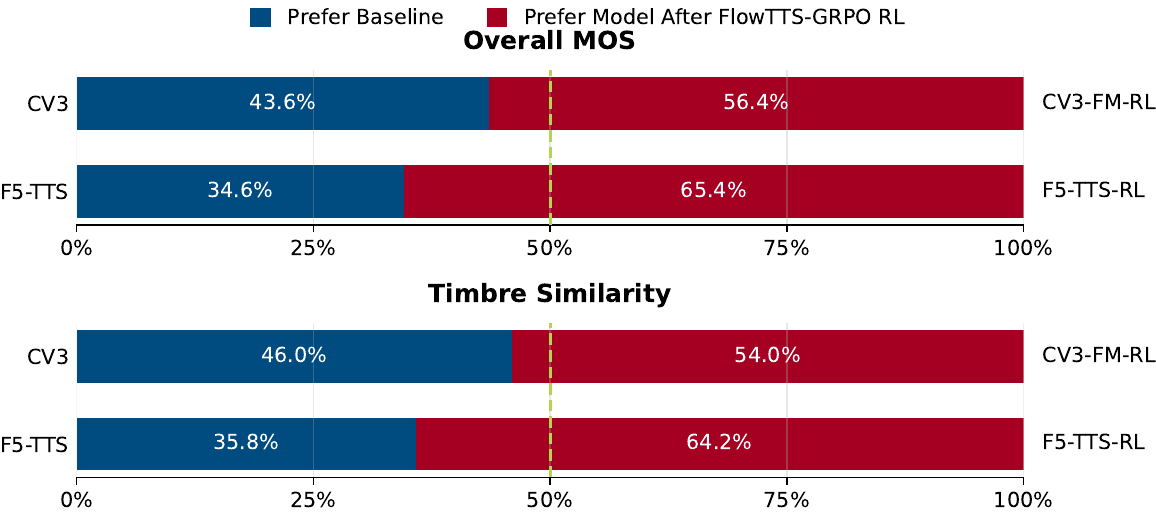}
	\caption{
        Subjective A/B preference test between the baseline and our FlowTTS-GRPO model on the English subjective evaluation set.
	}
	\label{fig:perfer_en}
\end{figure}

\begin{figure}[!t]
	\centering
\includegraphics[width=1.0\linewidth]{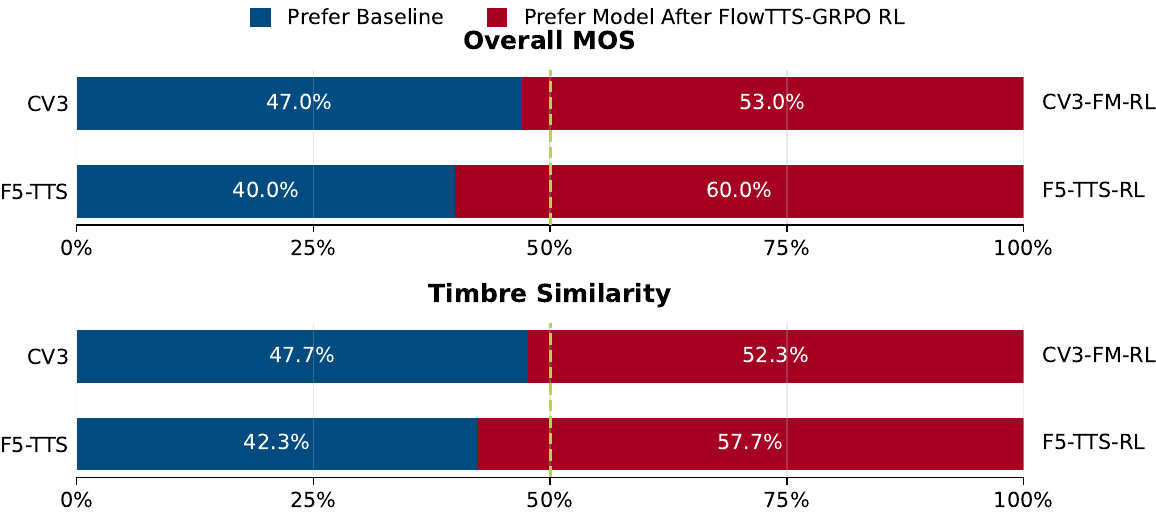}
	\caption{
        Subjective A/B preference test between the baseline and our FlowTTS-GRPO model on the Chinese subjective evaluation set.
	}
	\label{fig:perfer_zh}
\end{figure}

\subsection{Subjective Evaluation}
\label{sec:subject_eval}


To evaluate whether the objective reward gains translate to human perception, we conduct subjective A/B preference tests comparing F5-TTS, CV3, and their RL-finetuned counterparts (F5-TTS-RL and CV3-FM-RL). For each language, 30 samples are randomly selected from the evaluation sets for zero-shot synthesis, and 10 native speakers are recruited to perform paired comparisons. As illustrated in Fig. \ref{fig:perfer_en} and \ref{fig:perfer_zh}, the models optimized via FlowTTS-GRPO are consistently preferred over their respective baselines across both English and Chinese datasets. The clear margins in Overall MOS and Timbre Similarity preferences confirm that our method bridges the gap between objective proxy rewards and human auditory preferences. These results demonstrate that FlowTTS-GRPO achieves better perceptual alignment, leading to superior naturalness and more accurate voice cloning.

\section{Conclusion}

We introduce FlowTTS-GRPO, the first application of Flow-GRPO to text-to-speech models. Our method enables direct fine-tuning of open-source FM-only and LLM-FM hybrid models by converting ODE trajectories to SDE paths. Our framework simplifies prior RL approaches by eliminating value networks, preference pairs, and token-to-reward models, and we demonstrate that weighted reward combination with standard deviation normalization yields faster and more stable convergence than probabilistic schemes. We identify three practical optimizations: omitting classifier-free guidance during training accelerates convergence, hard case synthesis enhances robustness, and applying RL to the FM component primarily improves audio-detail metrics whereas RL on LLM targets intelligibility. Experiments on CosyVoice 3.0 and F5-TTS demonstrate significant improvements in speaker similarity and perceptual quality, with F5-TTS also achieving better intelligibility. Notably, FlowTTS-GRPO achieves state-of-the-art speaker similarity on Seed-TTS-Eval-zh, surpassing the closed-source Seed-TTS for the first time, and demonstrates robust generalization across speaker verification models, languages, and LLM front-ends.



\section{Generative AI Use Disclosure}

We use ChatGPT (OpenAI) and Qwen3-Max (Alibaba) for English grammar checking, sentence polishing in the manuscript. The AI tools are used only for grammar checking and did not generate any significant part of the scientific content, technical contributions, or experimental results. All authors are fully responsible for the content of this paper.

\bibliographystyle{IEEEtran}
\bibliography{mybib}


\end{document}